\documentclass{mn2e}
\usepackage{graphicx}  
\usepackage{epsf}
\usepackage{amsmath}
\newcommand{\hangpar}{\noindent\hangindent.1in}
\newcommand{\teff}[1]{$T_{\rm eff}$}
\newcommand{\vsini}[1]{$v\cdot\sin(i)$}
\def\cm1{$\rm cm^{-1}$}
\def\kms{$\rm km\,s^{-1}$}
\def\DE{D\kern-0.75em \raisebox{1.0pt}{=}\ }

\def\Sum{N_{\rm tot}}
\def\I{{\sc i}}
\def\II{{\sc ii}}
\def\Rv{\rule[-0.1in]{0.0pt}{20.0pt}}

\title[A detailed abundance analysis of 
HD 104237]
{The Herbig Ae SB2 System HD 104237
\thanks{Based on observations
obtained at the European Southern
Observatory, Paranal and La Silla, Chile
(ESO programmes 266.D-5655(A), 60 A-9036(A),
and 085.D-0296}}
\author[C. R. Cowley, F. Castelli,
S. Hubrig]
{C. R. Cowley${^1}$
\thanks{E-mail: cowley@umich.edu},
F. Castelli${^2}$, 
S. Hubrig${^3}$,
 \\
$^{1}$Department of Astronomy, University of Michigan,
   Ann Arbor, MI 48109-1042, USA\\
$^{2}$Istituto Nazionale di Astrofisica, 
Osservatorio Astronomico di Trieste, 
via Tiepolo 11, 34143, Trieste, Italy \\
$^{3}$Leibniz-Institut f\"{u}r Astrophysik Potsdam (AIP), 
An der Sternwarte 16, 14482, Potsdam, Germany  \\
}

\begin{document}

\date{Accepted . Received ; in original form }

\pagerange{\pageref{firstpage}--
\pageref{lastpage}} \pubyear{2013}

\maketitle

\label{firstpage}

\begin{abstract}
 
The double-lined spectroscopic binary HD 104237 (DX Cha) is part of a complex system of some half-dozen nearby young stars.  We report a significant change from an orbit for the SB2 system derived from 1999-2000 observations. We obtain abundances from the primary and secondary spectra.  The abundance analysis uses both detailed spectral synthesis and determinations based on equivalent widths of weak absorption lines with $W_\lambda$ typically $<$ 25\,m\AA.  Abundances are derived for 25 elements in the primary, and 17 elements in the secondary. Apart from lithium and zirconium, abundances do not depart significantly from solar. Lithium may be marginally enhanced with respect to the meteoritic value in the primary.  It somewhat depleted in the secondary. The emission-line spectrum is typical of Herbig Ae stars.  We compare and contrast the spectra of the HD 104237 primary and two other Herbig Ae stars with low $v\cdot\sin(i)$, HD 101412 and HD 190073.
\end{abstract}

\section{Introduction}
\label{sec:intro}

The star HD 104237 (DX Cha) is the brightest of the young Herbig Ae stars.  Its spatial relationship to several families of young stars is nicely illustrated by Feigelson, Lawson, and Garmire (2003, see their Figure 1). A detailed description of this complex multiple system is available from Grady, et al. (2004).  One component is very close to the primary so the spectrum is of an SB2.

B\"{o}hm, et al. (2004) discuss radial velocities of this
system in a ``binary approximation,'' that is, neglecting
any influence of some half-dozen or more nearby stars.  
They derive an orbital period of 19.859 days, and an
eccentricity of 0.665.  The systemic velocity was
+13.94 \kms.

A marginal magnetic
field detection reported by Donati, et al. (1997) was
not confirmed by Wade, et al. (2007).  Hubrig, et al.
(in preparation) now confirm the presence of a weak field.
Further discussion
of magnetic fields in HD 104237 and other Herbig Ae
stars is given by Hubrig, et al. (2007, 2009), and
Wade, et al. (2007).  

HD 104237
was among the ``24 dusty (pre-)main-sequence stars"
whose chemical abundances were determined by Acke
and Waelkens (2004, henceforth, AW). B\"{o}hm, et al. 
(2004) and more recently, Fumel \& B\"{o}hm
(2012, henceforth FB) made detailed studies of
the pulsation, and used the Fe~\I--Fe~\II\, lines
to fix \teff\, and $\log(g)$.  The present abundance study
builds on the work of AW and FB.  A more detailed
discussion of the present and earlier work is presented
in Section ~\ref{sec:prev}.  We do not know of a 
previous lithium abundance determination in a Herbig
Ae star.

\section{Spectra}
\label{sec:spec}

This analysis is based on the spectrum of HD 104237 
available from the UVESPOP archive (Bagnulo, et al
2003).  The spectral resolution of the UVES instrument,
normally 80,000, is roughly double that of the
material used by AW and FB.  The signal to noise of
the UVESPOP spectra are typically between 300 and 500,
according to Bagnulo, et al.  Tests on regions that
{\it appeared} free of spectral features were rarely
higher than 200, and were typically between 100 and 200. 

Additional material from the ESO archives include two HARPS spectra, which we designate (a) and (b). HARPS spectra have resolution 120000  (Mayor, et al. 2003) but the signal-to-noise was significantly lower than that of the UVESPOP spectrum.  We measured values between 50 and 60 for HARPS(a).  HARPS(b) has S/N 126 in the region of the Li \I\, line, but below 70 in tested regions short of 5200 \AA. Therefore, apart from work with the Li \I\, $\lambda$6707 feature (Sec.~\ref{sec:lithium}), all synthetic fits and equivalent width measurements were made on the UVESPOP spectrum. 
 
Observational epochs
are given in Table~\ref{tab:epoch}.  
We also give the heliocentric radial velocity of the
primary, and the difference (Primary minus Secondary) of
the radial velocities of both components.  Errors
are based on multiple measurements of several features.

The radial velocities of the primary star on 23 January 2003 and 13 January 2007 are higher than any value measured by B\"{o}hm, et al., or expected from their orbital parameters: $V_R(max) \approx +23$ \kms.  A significant change in the orbit has occurred since the earlier observations of 1999 and 2000.
\begin{center}
\begin{table}
\caption{Observations\label{tab:epoch}}
\begin{tabular}{l c c c} \hline
Spectrum & epoch &$V_R$ \kms &$\Delta V_R$ \kms  \\
UVESPOP & 23 January 2003 & $+36.7\pm 0.5$&$-20.4\pm 1.9$ \\
HARPS(a)& 13 January 2007&$+30.2\pm 1.$&$-19.1\pm 2.1$  \\
HARPS(b)& 03 May 2010    &$-0.6\pm 1.$ &$+37.8\pm 1.8$ \\  \hline
\end{tabular}
\end{table}
\end{center}

\section{Model atmospheres and spectral calculations}

The abundance analysis was performed by using both spectrum synthesis and equivalent widths. In the first case, model atmospheres for HD 104237 (primary P and secondary S) were computed with the ATLAS9 code (Castelli and Kurucz 2003); in the second case, the $T$--$\tau$ relations from the ATLAS9 models (Castelli 2010)
were used to regenerate the depth-dependences based on slightly different opacities and partition functions (Cowley, Adelman, and Bord 2003). Comparison of the two kinds of calculations showed negligible differences in computed features.
\section{Atmospheric Parameters}

The composite nature of the spectrum of HD 104237
precludes the use of broad or intermediate-band
photometry to fix $T_{\rm eff}$  and $\log(g)$.  
Therefore we based the parameter determination only on 
the high resolution spectra.

\subsection{Spectral synthesis}

Synthesized spectra of both stars were computed with
the SYNTHE code (Kurucz 2005).  These were added
to create a synthetic, composite spectrum.
This was then overplotted with 
the observed spectrum, and appropriate adjustments
to the stellar parameters were made to achieve 
optimum agreement. 

Atomic line lists are updated versions of those used by 
Castelli \& Hubrig (2004). They are mostly based on
the line lists available at the Kurucz (2012) 
website and on the NIST (2012) data base. 

 The computed spectra for both stars were  broadened for
a gaussian instrumental profile corresponding to a resolving power of 80000
for the UVES spectrum and of 120000 for the HARPS spectra.

\subsubsection{The primary\label{sec:prim}}

\begin{figure}
\includegraphics[width=84mm,height=60mm,angle=00]{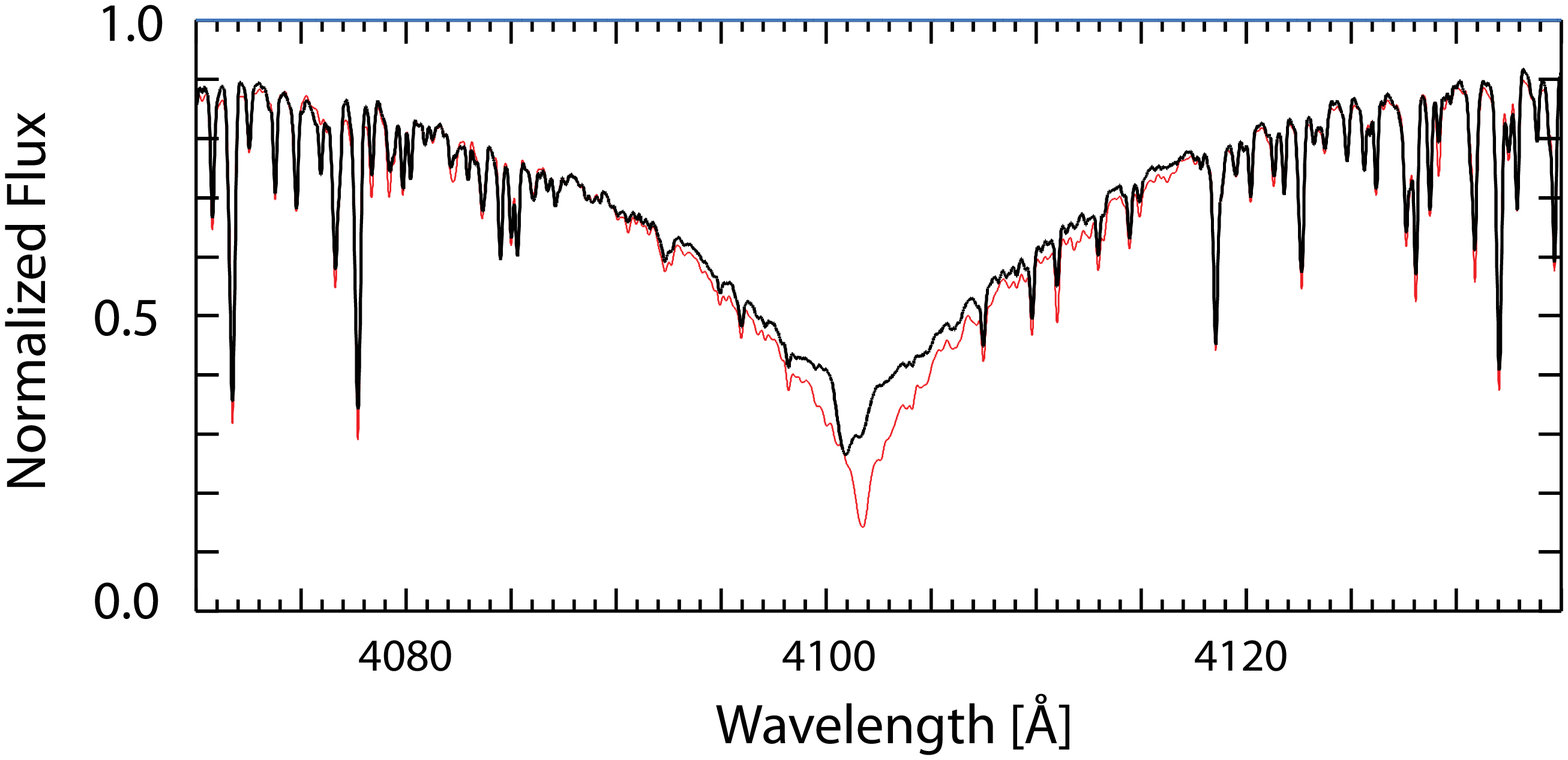}
 \caption{Synthesized H$\delta$ profile of the composite
spectrum (gray, or red online).  The observed spectrum
(black) shows a partially filled in, and shifted core.
\label{fig:hdel}}
\end{figure}

The derived atmospheric parameters of the primary star
$T_{\rm eff}$\,=\,8250\,K, and 
$\log(g)$\,=\,4.2 were based mostly on a
comparison of the observed and computed wings of H$\delta$, which is a
Balmer profile in HD\,104237 minimally affected by emission
(Figure~\ref{fig:hdel}).
These parameters were later confirmed by the agreement of the Fe\,\I\, and Fe\,\II\, abundances obtained from both spectrum synthesis and equivalent widths (see below).

Because the strong profiles of the primary star were not well reproduced
after broadening by a rotational profile, we added a gaussian macroturbulent velocity.
The adopted values were  
$v\cdot\sin(i)$ of 8\,\kms\, for the rotational velocity
and $\xi_M = 9$\,\kms\, for the gaussian macroturbulent velocity. 
\begin{figure}
\includegraphics[width=90mm,height=85mm,angle=90]{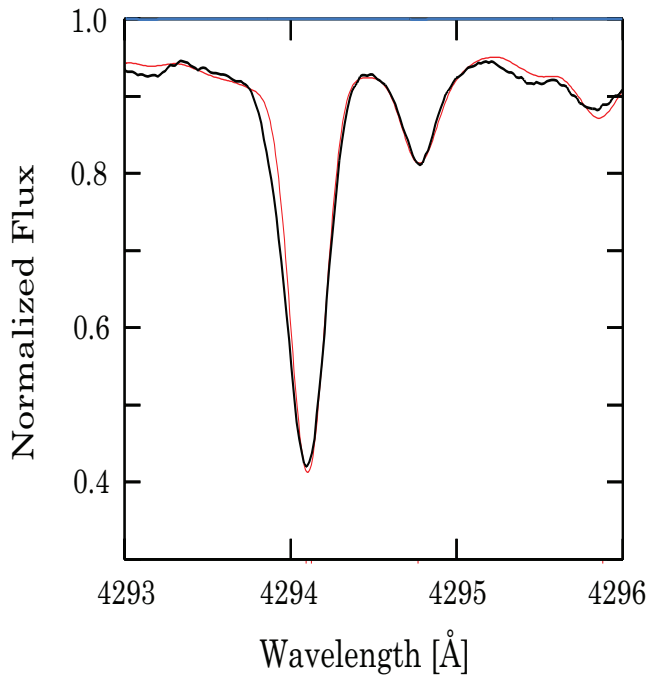}
\caption{Comparison of the asymmetric observed profile at 4294.1\,\AA\ 
(black line) with the synthetic blend profile due to Ti\,II 4294.094\,\AA\ 
and Fe\,I 4294.125\,\AA\ (gray, red online).\label{fig:asymm}}
\end{figure}

While the weak and medium-strong profiles were well 
predicted by the combined velocity broadenings, the 
strong lines showed an asymmetric profile with a more 
extended violet wing that was impossible to reproduce 
with our computations. 
An example is shown in Figure~\ref{fig:asymm}.
This kind of profile which could indicate an expanding 
atmosphere, is seen in many lines in both the UVES and 
HARPS spectra.

The microturbulent velocity value
$\xi_t=2.5$\,\kms\,  for the primary 
star was determined both from
the equivalent widths and the synthetic spectrum methods.

\subsubsection{The secondary\label{sec:sec}}

The spectrum of HD\,104237 is contaminated by that of the close companion.
FB analyzed a spectrum that was obtained by subtracting a  
computed spectrum for the secondary from the observed spectrum.

We have preferred to analyze the observed spectrum as it is, and have tried to obtain information also from the secondary spectrum. This was not trivial, mostly because no usual spectroscopic methods for the parameter determinations can be applied to the close companion. In fact, neither hydrogen lines nor elements with lines in two different ionization stages are independently observable. We based our parameters  for the secondary mostly on the Ca I line  at  6122.21\,\AA\ which is a good luminosity indicator (Cayrel et al. 1996). The two components for this line are rather well separated  and free of blends. Unfortunately, the other two lines of the Ca I triplet at 6102.73\,\AA\ and 6161.30\,\AA\ are blended.

The composite synthetic spectrum normalized to the continuum (PS) was computed  according to the relation (Lyubimkov \& Samedov, 1987)\\

\begin{equation}
\frac{F_{\lambda}(PS)}{F_{\lambda c}(PS)}=\frac{F_{\lambda}(P)R_P^2+F_{\lambda}(S)R_S^2}
 {F_{\lambda c}(P)R_P^2+F_{\lambda c}(S)R_S^2}. 
\label{eq:lyubim}
\end{equation}
\noindent We use the subscript `$\lambda c$' to indicate the
continuum $at$ a wavelength $\lambda$.  The $F$'s for the
primary (P) and secondary (S) have units of energy per
unit area, per sec, per {\AA}ngstr\"{o}m.
The $R$'s, are the stellar radii.


After some experimentation with alternate models and different radii ratios, we selected a ratio \( R_S/R_P=1 \) (see Section~\ref{sec:consis}), and the parameters \teff\, =\,4800\,K and log(g) = 3.7 for the secondary. A rotational velocity v\,sin(i) = 12\,km\,s$^{-1}$ was deduced from the comparison of the observed and computed  profiles.  We thank the referee, Barry Smalley, who pointed out that the same v\,sin(i) and macroturbulence (8 and 9 \kms, respectively) would have nearly the same combined profile as a v\,sin(i) of 12 and no macroturbulence.  This would be the case if the stars were tidally locked.

A microturbulent velocity $\xi_t$\,=\,1.0\,km\,s$^{-1}$ and solar abundances were assumed. FB used \teff\, =\,4500K, log(g)\,=\,4.0 and assumed solar abundances and a microturbulent velocity $\xi_t$\,=\,2\,km\,s$^{-1}$.

\begin{figure}
\includegraphics[width=85mm,height=84mm,angle=90]{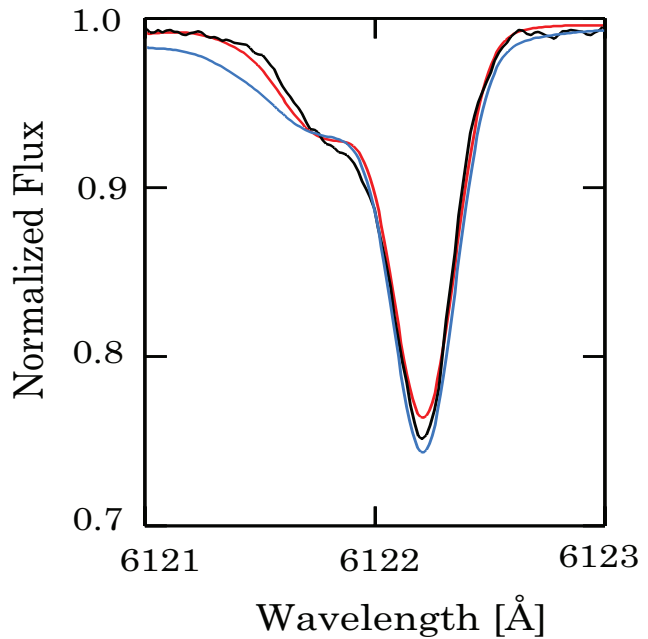}
 \caption{Observed (black) and calculated Ca \I\, $\lambda$6122.21, with $\log(g) = 3.7$ (light gray/red online), and $\log(g) = 4.0$ (dark gray/blue online).   The main feature is due to the primary.  Absorption from the same line in the secondary star (see text) is to the violet. \label{fig:ca1}}
\end{figure}

Figure~\ref{fig:ca1} compares the observed Ca I profile at 6122.21\,\AA\ with the 
composite profiles computed for $T_{\rm eff}$\,=\,8250\,K, and 
$\log(g)$ = 4.2 for the primary and for both $T_{\rm eff}$ = 4800\,K, and 
$\log(g)$ = 3.7 and $T_{\rm eff}$ = 4500\,K, and  $\log(g)$ = 4.0 for the secondary.
In the first case the ratio of radii is \( R_S/R_P = 1 \), in the second case
it was deduced from the relation 
\begin{equation}
\frac{R^2_S}{R^2_P} = \frac{M_S}{M_P}\cdot\frac{g_P}{g_S}.
\label{eq:rradg}
\end{equation}
 for a mass ratio M$_{P}$/M$_{S}$\,=\,1.29
(B\"ohm et al. 2004). 
The too-broad wing of the secondary 
star profile computed for log($g$)\,=\,4.0 is evident. 

The adopted parameters for the two components (Table.~\ref{tab:adpar}) were well suited 
to reproduce
several  primary-secondary pairs of profiles as Fe\,\I\, at 4950.11, 4969.92, 5002.793, 
5393.17, 5576.09\,\AA, 6065.48, 6252.56, 6335.33\,\AA, 
Ca\,\I\, 6717.68\,\AA, and several others.


\subsection{Equivalent widths}

\begin{figure}
\includegraphics[width=70mm,height=80mm,angle=-90]{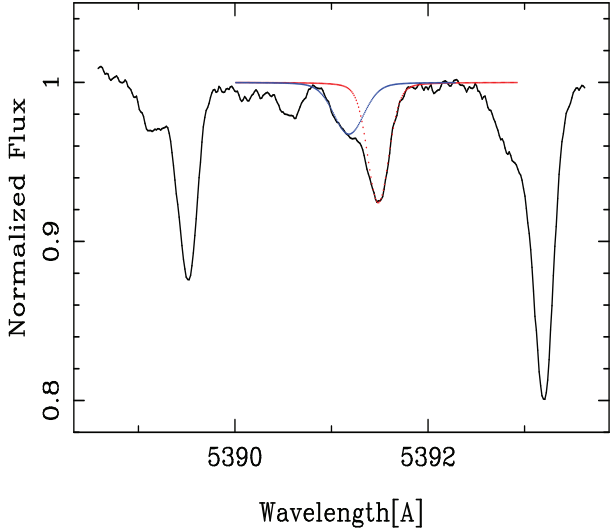}
 \caption{Fe \I\, $\lambda$5391.46, showing Voigt-profile 
fits to absorptions from components P and S (weaker, violet
component).  
Lines from the secondary can also be seen in the stronger
lines, Fe \I, $\lambda\lambda$5389.49, and 5393.17
\label{fig:sb2}}
\end{figure}

A closely related approach makes use of measurements
in the observed  composite spectrum 
(Figure~\ref{fig:sb2}).  Voigt-profile fits are 
used to derive equivalent widths.
These values are then related to computed equivalent 
widths from the adopted models, assuming solar 
abundances.  

Equivalent widths were not used to derive
atmospheric parameters or abundances for the secondary
star, with the exception of lithium (Sec.~\ref{sec:lithium}).

Let $x_\lambda = L_{\lambda c}(S)/L_{\lambda c}(P)$ 
be the {\it continuum} luminosity ratios, where
$L_{\lambda c}(S) = F_{\lambda c}(S)\cdot 
4\pi R_S^2$, and similarly
for the primary.  
With the parameters given in Table~\ref{tab:adpar},
the flux ratio $F_{\lambda c}(S)/F_{\lambda c}(P)$ at 
$\lambda$\,=\,5576\,\AA\ is $1/9.9$.  Because of our
assumption of equal stellar radii, the ratio is
the same for the luminosities.

Directly measured equivalent widths from the composite
spectrum must be corrected to account for the relative
contribution of the light of the two stars.  
Assume that the flux 
in a line of the primary component is diluted only by the 
continuum flux of the secondary component.  Then the
equivalent width of a line of the P-component
measured in the composite spectrum, $W_P^{PS}$,
is related to the equivalent width, $W_P$ in the
undiluted spectrum by 
\begin{equation}
W_P = (1+x_\lambda)\cdot W_P^{PS}
\label{eq:WWA}
\end{equation}

Using a similar notation, the (undiluted) equivalent 
width of the secondary component may be obtained from
\begin{equation}
W_S = \frac{1+x_\lambda}{x_\lambda}\cdot W^{PS}_S.
\label{eq:WWB}
\end{equation}



Figure~\ref{fig:sb2} illustrates measurements on the composite
UVESPOP spectrum.  It can be seen that the assumption
that the lines of the two components are separate from
one another is only approximately true.  This assumption
is implicit in Eqs.~\ref{eq:WWA} and ~\ref{eq:WWB}.
However, direct
comparisons between equivalent width and results from
complete synthesis (including the secondary), are in
good agreement, showing that the approximation used
here are
good to a few hundredths of a dex.  At longer wavelengths,
the separation of the components is greater than that
shown in Figure~\ref{fig:sb2}.

Equivalent widths of Fe \I\ and Fe \II\ lines were measured,
and corrected as indicated.  This is basically an iterative
procedure. The fraction $x_\lambda$ in Eqs.~\ref{eq:WWA} and
~\ref{eq:WWB} follows from measurements of the diluted
($W_P^{PS}$, and $W_S^{PS}$) and $calculations$ of the 
undiluted ($W_P$ and $W_S$) equivalent widths.  Calculations
of the latter require models, but initial models are 
available from FB.  Adopted models must give similar values
of $x_\lambda$ from either Eq.~\ref{eq:WWA} or ~\ref{eq:WWB}.

If the flux ratios, $x_\lambda$, are known, equivalent widths
of Fe \I\, and Fe \II\, lines may be used along with a grid of
atmospheres to determine a relation between \teff\,
and $\log(g)$.  

We also examined line-by-line abundances for other atomic and
ionic species with numerous lines, to check that the
abundances did not drift with excitation potential.  This
provides an independent check on \teff.

The overall procedure has the advantage that
weak lines ($W_\lambda \approx$ 25\,m\AA\, or less) may be
used.  These are in principle, independent of line
broadening mechanisms and the instrumental profile.

The synthesized spectra were useful for the selection of
lines that were free from blends, and could be used for 
abundances in the equivalent width method.  Indeed, the synthetic
spectrum automatically includes relevant line absorption from the
secondary when it overlaps with lines of the primary star.
Contributions from
the secondary are indicated automatically on the synthesized
spectrum,
making it straightforward to see when a feature is
influenced by an accidental overlap with an absorption from
the secondary star.  


The adopted models made use of both spectrum synthesis and
equivalent widths.  Their parameters are given in 
Table~\ref{tab:adpar}.   Uncertainty estimates for the primary are based on comparisons of results of calculations based on a grid of models with parameters
surrounding the adopted values.  

The comparison of the observed profiles of Ca I at 6122 A with profiles computed for several different choices of Teff and logg has lead us to
select as best parameters \teff = 4800 K and log($g$) $\le$3.7,
with an uncertainty of 0.2 dex.
While there is an uncertainty of about 200K for the temperature, the value of the gravity is an upper limit.  In fact, values of log($g$)= 3.6 and 3.5 would have been  as well suited to reproduce the observed profile for
temperatures differing within $\pm$200K  from 4800K.

AW used \teff\, =\, 8000K, and $\log(g) = 4.5$.  FB's
values were 8550K, and $\log(g) = 3.9$, for the primary.

\begin{table}
\caption{Adopted parameters for primary and secondary
atmospheres\label{tab:adpar}}
\begin{tabular}{l c c c} \hline
       &$T_{\rm eff}$K &$\log(g)$&$\xi_{\rm turb}$\\ \hline
Primary&  8250$\pm$200&4.2$\pm$0.25 &2.5$\pm$1  \\
Secondary&4800$\pm$200&$3.5 -\le 3.7$ &1.0$\pm$1 \\   \hline
\end{tabular}
\end{table}

\subsection{Pulsational studies\label{sec:seis}}

Astroseismological measurements are being used to predict
masses, radii, luminosity, and ages of stars (Mathur, et al. 2012).
Dupret, et al. (2006) attempted to fit the pulsation 
frequencies discussed by B\"{o}hm, et al. (2004) to 
theoretical models.  They 
found that the stellar properties that fit the 
observed frequencies were not in good agreement with 
best estimates of these values by other methods.  
Subsequently, Dupret et al. (2007) 
suggested He accumulation in the partial ionization
zone to account for the observed modes.

While FB reported extensive pulsational observations
and analysis,
they chose classical spectroscopic methods
to determine \teff\, and $\log(g)$ for HD 104237.  Those
methods are basically 
the same as the ones used in the present work.  We
conclude that astroseismic constraints need to be 
complemented by spectroscopic analysis; pulsational
properties alone are not yet sufficient, at least
for HD 104237.

\subsection{Radii, masses, and consistency\label{sec:consis}}

One approach to the ratio of the radii of the primary
and secondary of the HD 104237 system is to use the
ratio of absorption lines, such as those shown in Fig~\ref{fig:sb2}.
We measured equivalent widths of five Fe \I\, lines 
$\lambda\lambda$ 6065.48, 6252.56, 6336.82, 6411.65
\& 6421.35, to derive the factor $x_\lambda = 0.155$ of 
Eqs. ~\ref{eq:WWA} and ~\ref{eq:WWB}.  Since that 
ratio is defined by 

\begin{equation}
x_\lambda =\frac{L_{\lambda c}(S)}{L_{\lambda c}(P)}
=\frac{R^2_S}{R^2_P}\cdot\frac{F_{\lambda c}(S)}{F_{\lambda c}(P)}
\label{eq:fixrad}
\end{equation}
we may use the empirically determined $x_\lambda$ to
obtain the ratio of the radii.  Note the luminosities and
fluxes are wavelength dependent.
The models of Table~\ref{tab:adpar} provide the surface
fluxes for each of the stars.  That ratio at 6310\,\AA\ is 0.139, for
the surface flux of the secondary to that of the primary.
With these numbers  we obtain
$R_S/R_P = 1.06$, or unity within the uncertainties
discussed in this section and given in Table~\ref{tab:adpar}.

Another way for finding the ratio of radii is the use of the
Stefan-Boltzmann law. Here, we need the total luminosities
(integrated over wavelength).  Then
\begin{equation}
\frac{L(P)}{L(S)}=\frac{R^2_P}{R^2_S}\cdot\frac{T^4_{eff}(P)}{T^4_{eff}(S)}
\end{equation}
Assuming a luminosity ratio of 10 (B\"ohm et al. 2004)
the ratios of radii  for
T$_{eff}$(P)\,=\,8250\,K and T$_{eff}$(S)\,=\,4800\,K 
is R$_{P}$/R$_{S}$\,=\,1.07.

Yet another approach to finding the ratio of the radii is
to use the spectroscopically determined surface 
gravities and Eq.~\ref{eq:rradg}.
\noindent 
Assuming a mass ratio $M_P/M_S$ of 1.29  (B\"ohm et al. 2004), 
$\log(g_S) = 3.7$, and $\log(g_P) = 4.2$, the ratio of 
radii R$_{P}$/R$_{S}$ is 1.6. For R$_{P}$/R$_{S}$\,=\,1, and the same gravities 
as before, we obtain $M_P/M_S$\,=\,3.15.

Therefore, while we found a rather good consistency between the ratios of
radii, luminosities, and effective temperatures related by the 
Stefan-Boltzmann law, we obtain a manifest inconsistency between the ratios 
of radii, masses, and gravities related by the universal gravitation  law.
The mass ratio obtained by us is in 
conflict with the value of 1.29 of B\"{o}hm, et al. (2004). However, we
must take the uncertainties into account.

We note that the spectroscopic gravity for both the primary and secondary stars
are affected by errors which depend  first of all on the position of 
the continuum, which is rather difficult to draw, in particular above 
the hydrogen lines.
Furthermore, the parameters for the secondary are based on 
too many assumed or estimated quantities to allow 
quantitative analyses of the secondary 
unaffected by some large uncertainties.

Spectroscopic gravities typically
have an uncertainty of 0.2 (or more) in the log.  If we were to
assume 4.0 and 3.9 for the surface gravities of the
primary and secondary, then, still with unity for the
ratio of the radii, we would obtain 1.26 for the
mass ratio.  These gravity changes would affect the
derived abundances by only a few hundredths of a dex.
B\"{o}hm, et al. (2004) assign an uncertainty of
0.02 to their mass ratio, so their value could be
1.27.  While there is room for future adjustments
of the fundamental parameters of this system, we 
conclude there is no significant conflict of our 
parameters with the spectroscopic mass ratio.


\section{Abundances}
\label{sec:abundances}
\subsection{Previous abundance work\label{sec:prev}}

In this section, we briefly compare and contrast our
work with that of AW and FB.  In the latter study, only
the iron abundance was determined--independently from
Fe~\I\, and Fe~\II.

Both AW and FB
study derive abundances from 
equivalent widths.
Although AW did not account for 
dilution by the secondary spectrum, their abundance
results are similar to ours.  For the well-represented 
iron spectra (Fe~\I\, and ~\II\,) differences are under
0.1 dex.  
Differences are less than a factor of two,
apart from Cu, and two trace species, Nd and Sm, where AW 
report rather large excesses of +0.59 and +0.85 dex.
If we average the absolute value of the logarithmic 
differences, ``This study minus AW'', 
the value is 0.13 dex
for the elements through barium. 

FB (see their Table 5) give $-$4.38 for log(A$_{\rm Fe}$), 
or $\log(Fe/H)$.  This becomes $-$4.42 for $\log(Fe/N_{tot})$.
This differs from our value (Table~\ref{tab:abtab})
by only 0.02 dex.

\begin{figure}
\includegraphics[width=84mm,height=84mm,angle=0]{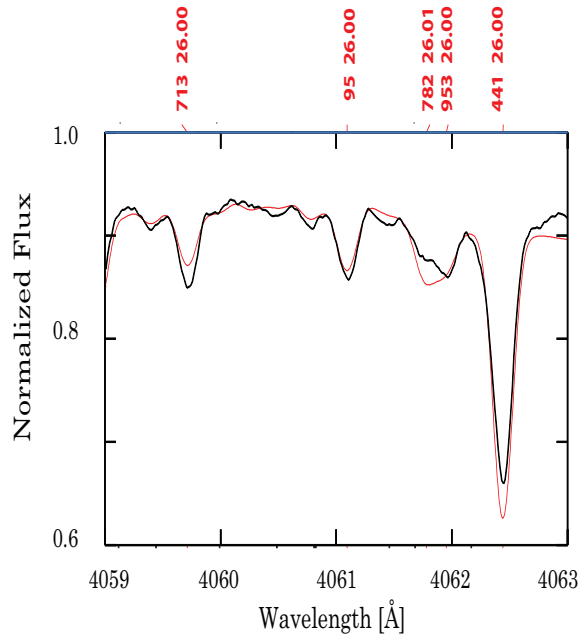}
 \caption{A previously unclassified Fe I line
at $\lambda$4061.09
(34362.87--58979.82 cm$^{-1}$) very closely coincides with 
a strong line of Nd II.  Synthesis shows the corresponding
feature in HD 104237 is mostly Fe I.  A minor contribution
of a few m\AA, commensurate with a solar Nd abundance
is likely.  Labels at the top of the plot indicate lines
from the primary (gray/red online).  
The decimal fraction of the 
wavelength is followed by the element (e.g. 26.01 is
Fe \II). The standard SYNTHE plots also contain information on line
strengths, and excitation. \label{fig:faknd}}
\end{figure}

Lines from the secondary and weak, previously unclassified
lines could account for the large excesses of Nd and Sm
reported by AW.
An example is shown in Figure~\ref{fig:faknd}, where
Fe I $\lambda$4061.09 coincides with Nd II $\lambda$4061.09.
The upper
level of the Fe I line in question was given by Nave, et
al. (1994), but is not in Sugar and Corliss (1985).

FB used abundances from
Fe~\I\, and Fe~\II\, to fix \teff\, and $\log(g)$.
Both FB and AW used excitation to break the 
\teff--$\log(g)$\, degeneracy.  Neither work computed
Balmer profiles.

With the help of estimates of the \teff\,
and $\log(g)$ of the secondary, FB synthesized its
spectrum, and subtracted it from the observed 
composite.  This method therefore differed from 
the present technique, in which a composite spectrum
was synthesized and compared directly with the observed
spectrum.  We see no particular advantage or disadvantage
of subtracting the secondary rather than synthesizing 
the composite spectrum.  While FB took the secondary into
account in their analysis of the primary, they do not
report abundances for the secondary.

Given an accurate synthesis of the
secondary, FB's method assured their equivalent widths
were unperturbed by blends from the secondary.  In the
our equivalent width method, such blends were avoided with the 
help of the results from the composite synthesis, where
contributions from the primary and secondary spectrum are
indicated on the plots (see Figure~\ref{fig:faknd}).

FB were careful to make use of lines with well-defined
continua, and accurate oscillator strengths.  These
constraints, however, reduced the number of usable 
lines to 8 Fe~\I\, lines and only 2 Fe~\II\, lines.
Just 3 of these lines were below 30 m\AA.  The
remainder were subject to saturation corrections
dependent on the microturbulence, $\xi_t$.  This was not 
independently determined, but fixed at 2 \kms, based
on previous studies.  AW used $\xi_t = 3$ \kms.

Table~\ref{tab:micro} illustrates the influence of the
assumed microturbulence on abundance for two Fe~\I\,
lines, $\lambda$4736 (74\,m\AA) and $\lambda$4485 (48\,m\AA).
The entries give the difference in logarithmic abundance
from that obtained with $\xi_t = 0.0$.  We see that for
the 74\,m\AA\, line, the abundance difference using 
2.0 and 2.5 \kms\, is 0.82 $-$ 0.69 = 0.13 dex.  For the
weaker line, the difference is only 0.02 dex. For 
reference, note that in a recent study of Herbig AeBe 
stars, Folsom, et al. (2012) reported values of $\xi_t$ 
between 1.3 and 3.7 \kms.
 \begin{center}
\begin{table}
\caption{The effect of microturbulence on abundance
\label{tab:micro}}
\begin{tabular}{c c c} \hline

$\xi_t$(\kms)  &\multicolumn{2}{c}{log(diff)} \\
             &4736(74m\AA)&4485(48m\AA) \\ \hline
0.0          & 0.00       & 0.00         \\
1.0          & 0.26       & 0.04          \\         
2.0          & 0.69       & 0.09   \\
2.5          & 0.82       & 0.11   \\
3.0          & 0.91       & 0.13  \\
3.5          & 0.98       & 0.14  \\  \hline
\end{tabular}
\end{table}
\end{center}
We determined $\xi_t$ independently for each spectrum,
obtaining values ranging up to 3.3 \kms.  Eventually
we adopted 2.5 \kms, which was more compatible with
the spectrum synthesis fits.  However, our abundances
based on equivalent widths of weak lines are 
virtually independent of this choice.

\subsection{Present results\label{sec:results}}
 Equivalent widths, oscillator strengths, and 
abundances are available in the online material
for some 375 absorption lines.
Abundances were obtained from these
equivalent widths using the Michigan
software, usually for for the weak lines and 
are in consistent agreement with 
best fits from detailed spectrum synthesis.

Abundances are summarized in Table~\ref{tab:abtab}.
Some spectra, e.g. Ca \II, 
were not used for abundances, even though their
lines were clearly present.  This was 
usually because of
blends and/or oscillator strengths with low accuracy.
Such cases are indicated
by ``n.u.'' (not used), in the table.
Probable errors (pe) are standard deviations of abundances
from the number of lines given in the column labeled `$n$'.
For $n = 2$, the error is the difference of
abundances.  For an element, e.g. Fe (Fe \I\, and
Fe \II), the error is the difference of values from the
atomic and ionic spectra.
The column marked `AW' gives logarithmic abundances computed
from AW's paper.  We used the Anders \& Grevesse (1998) 
solar abundances
to compute $\log(El/N_{tot})$ from the bracket values 
given by AW.  One exception is Fe, where AW used 7.52
(log(H) = 12) rather than 7.67 given by Anders and Grevesse
for the photospheric iron abundance.
The solar abundances in Column 7 are from
Asplund, et al.(2009).
\begin{table}
\caption{Adopted abundances.  Column 2 is a key to
oscillator-strength references given in the appendix.
See text for a description of the column labeled AW.
\label{tab:abtab}}
\begin{minipage}{17cm}
\begin{tabular}{l c r c r r r} \hline
Spec & $gf$-ref& $\log(El/\Sum)$&pe    &n&AW &Sun \\ \hline
Li {\sc i} &1   &Table~\ref{tab:li1} & \\
C {\sc i} &2   & $-$3.64 & 0.12  & 8&$-$3.62 &$-$3.61\\
N {\sc i} &3   & $-$4.13 & 0.07  & 4&$-$4.41 &$-$4.21 \\
O {\sc i} &4   & $-$3.31 & 0.09  & 5&$-$3.21 &$-$3.35\\
Na {\sc i}&5   & $-$5.64 & 0.01  & 2&$-$5.64 &$-$5.80\\
Mg {\sc i}&6   & $-$4.79 & 0.30  & 5&$-$4.58 &$-$4.44\\
Mg {\sc ii}&7  & n.u.    &       &   &$-$4.65     \\        
Al {\sc i}&8   & $-$5.66 & 0.30  & 4&$-$5.59&$-$5.59 \\
Si {\sc i}&9   & $-$4.51 & 0.44  & 7&$-$4.47&$-$4.52\\
Si {\sc ii}&10  & n.u.   &       &   &$-$4.18 \\
S {\sc i}  &11  & $-$4.70 & 0.02  &  2&$-$4.98& $-$4.92 \\
Ca {\sc i} &12  & $-$5.79 & 0.08  &  4&$-$5.53&$-$5.70\\
Ca {\sc ii}&13  & n.u.    &       &   &$-$5.54  \\
Sc {\sc ii}&14  & $-$8.63 & 0.20  &  5&$-$8.87& $-$8.98\\
Ti {\sc i} &15  & $-$7.05 & 0.11  &  3  &      \\
Ti {\sc ii}&16  & $-$7.12 & 0.16  &  5  &$-$6.82   \\
Ti         &    & $-$7.08 & 0.07  &    & &$-$7.09 \\
V {\sc i}  &17  & $-$8.06 & 0.15  &  1  &  \\
V {\sc ii} &18  & $-$7.77 & 0.21  &  3  &         \\
V          &    & $-$7.97 & 0.29  &     &$-$8.05 &$-$8.11  \\
Cr {\sc i} &19  & $-$6.31 & 0.25  & 7   & \\
Cr {\sc ii}&20  & $-$6.26 & 0.13  & 7   &$-$6.20  \\
Cr         &    & $-$6.29 & 0.05  &     & &$-$6.40 \\
Mn {\sc i} &21  & $-$6.63 & 0.17  & 4  &$-$6.70& $-$6.61\\
Mn {\sc ii}&22  & n.u.    &       &     &  &     \\
Fe {\sc i} &23  & $-$4.37 & 0.13  & 22  &$-$4.44          \\
Fe {\sc ii}&24  & $-$4.42 & 0.17  & 17  &$-$4.42     \\
Fe         &    & $-$4.40 & 0.05  &     & & $-$4.54  \\
Co {\sc i} &25  & $-$7.26 & 0.08  & 2  & &$-$7.05 \\
Ni {\sc i} &26  & $-$5.70 & 0.19  & 10  &$-$5.70& $-$5.82 \\
Ni {\sc ii}&27  & n.u.    &       &  2  &         \\
Cu {\sc i} &28  & $-$7.96 & 0.25  &  3  &$-$7.57& $-$7.84 \\
Zn {\sc i} &29  & $-$7.58 & 0.25  &  2  &$-$7.81& $-$7.48\\
Sr {\sc ii}&30  & $-$9.17:& 0.30: &  2  &       & $-$9.17\\
Y {\sc ii} &31  & $-$9.69 & 0.13  &  4  &$-$9.62& $-$9.83 \\
Zr {\sc ii}&32  & $-$9.06 & 0.06  & 4  &$-$9.09& $-$9.46 \\
Ba {\sc ii}&33  & $-$9.61 & 0.20  & 5  &$-$9.60& $-$9.86 \\ 
La {\sc ii}&34  & $-$10.72& 0.10 & 3  &$-$10.72 &$-$10.94 \\ \hline
\end{tabular}
\end{minipage}
\end{table}


Most elements are solar within the errors.  
We examined features near lines of the light lanthanides,
Ce-Sm, but they were weak and often blended with absorption
from the secondary.  We see no indication these elements are
significantly different from solar in abundance.

A case
could be made for slight enhancements of a few elements.
The case for Zr is the strongest.   There
are numerous lines, modern gf-values
(Ljung, et al. 2006) and the abundance excess is
several times the uncertainty.  The lithium
abundance is discussed below.


\subsection{The secondary}

In addition to Li\,I, we identified in the secondary lines of 
Na\,I, Mg\,I, Al\,I, Si\,I,  Ca\,I, Sc\,I, Ti\,I, V\,I,  Cr\,I, Mn\,I, Fe\,I, 
Co\,I,  Ni\,I, Cu\,I, Zn\,I, Sr\,I, Ba\,II. 
A large number of molecular lines, especially of C2, CN, CO, CH, MgH 
contribute to lower the continuum.
We found several underabundances.  Logarithmically, relative to
solar, they were for Sc [$-$0.3], Ti [$-$0.4], V [$-$0.4],
Cr [$-$0.4], Co [$-$0.7], Ni [$-$0.3]. Furthermore, in addition to Li,
we found an overabundance for Ba [+0.5]. All the other elements are
well predicted by solar abundances. 
We note that these abundances are estimates derived from
the comparison of the observed UVESPOP spectrum with the synthetic spectrum.
More work would be needed for an accurate quantitative analysis of the 
companion of HD\,104237. 

\subsection{Lithium\label{sec:lithium}}

While lithium is commonly observed in young, cool
stars, we are unaware of its observation in other Herbig
Ae stars.  Of course, a strong Li \I\, line 
was noted in HD 104237
by FB and others, and correctly attributed to the
cooler secondary.  B\"{o}hm, et al. (2004) discuss
the lithium line in both primary and secondary,
but we know of no prior abundance determinations
from either line in HD 104237.

\begin{figure}
\includegraphics[width=84mm,height=210mm,angle=00]{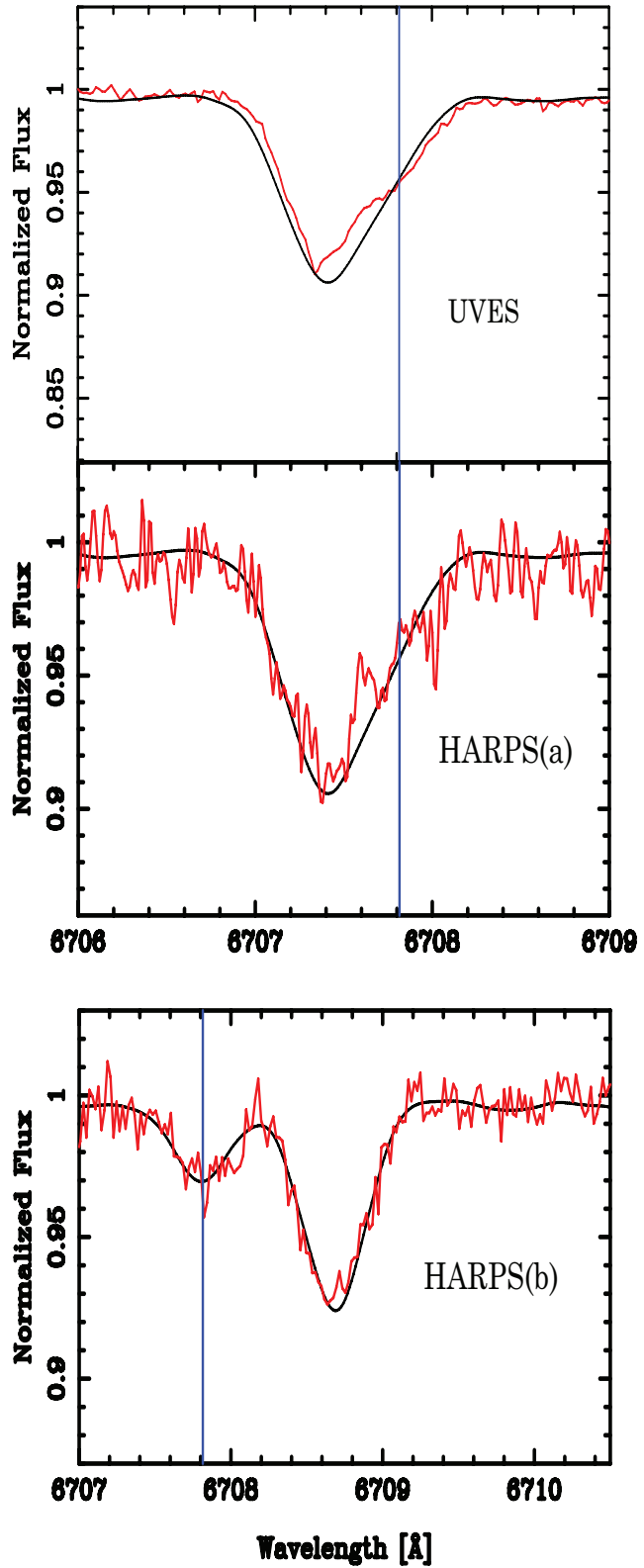}
 \caption{The Li \I\, region, showing synthetic fits.  
See Section ~\ref{sec:spec} for 
epochs of the UVES, HARPS(a) and HARPS(b) spectra
(gray, red online).
The vertical line marks the rest
wavelength of Li I in the primary.  Abundances 
in Table~\ref{tab:li1} were used for the lower plot.  
Note the different wavelength scales for the upper two,
and the lower plots.
\label{fig:li1}}
\end{figure}

Lithium is the only light element that shows a clear
departure from the solar {\it photospheric} abundance.
The Li \I\, region is shown
in Figure~\ref{fig:li1}. 
The weaker feature marked by a vertical line, is 
Li \I\, in the primary.  Note the stronger feature,
Li \I\, in the secondary, is to 
the red in HARPS(b).

An abundance is derived for both primary and secondary
spectra from equivalent widths as well as synthesis.
Only abundances from the HARPS(b) spectrum (Figure~\ref{fig:li1}, below), where the primary and secondary lines are
well separated, are adopted.
Note that when we apply the correction of Eq.~\ref{eq:WWB}
to the equivalent width for the secondary, 38.1 m\AA,
using $x_\lambda$ = 0.171, we get 261 m\AA, a line no
longer weak.  The abundance, $\log(Li/N_{tot}) = -9.28$
was calculated with the full hfs structure (cf. Smith,
Lambert \& Nisson 1998).  A solar mix of $^6$Li and
$^7$Li was assumed.

To estimate errors in the lithium abundances, we
use HARPS(b) values.
Comparing abundance results 
from from synthesis and equivalent widths, we estimate
the uncertainties for the lithium abundances are about
0.1 dex for the primary and 0.25 dex for the secondary.


The lithium abundance for the primary may be
slightly enhanced with respect to the
meteoritic value (Lodders, Palme \& Gail 2009).  There 
is evidence for some destruction
of lithium in the cooler secondary.

\begin{table}
\begin{center}
\caption{Lithium abundances 
from $\lambda$6707.  See text for uncertainty estimates.\label{tab:li1}}
\end{center}
\begin{tabular}{l c c c} \hline
\multicolumn{2}{c}{{\Rv}Primary}&
\multicolumn{2}{c}{Secondary} \\ \hline
{\Rv}$W_\lambda$/Rem &$\rm\log(Li/N_{tot})$&
$W_\lambda$&$\rm\log(Li/N_{tot})$ \\ \hline
13.6/HARPS(b)&$-$8.58  &38.1           &$-$9.28 \\
synthesis:   & \\
uves         &$-$8.64  &               &$-$8.90 \\ 
HARPS(a)     &$-$8.64   &              &$-$8.90 \\
HARPs(b)     &$-$8.64   &              &$-$9.50 \\
(solar photosphere:)&$-$10.99 \\
(meteorite:)  &$-$8.79  \\  \hline
\end{tabular}
\end{table}

\section{The emission-line spectrum\label{sec:emiss}}

In this section we compare and contrast the emission-line
spectra of HD 104237 with two sharp-lined Herbig Ae stars
HD 101412 and HD 190073.  Spectra and abundances for these
stars were discussed by Cowley, et al. (2010), and Cowley \& Hubrig (2012).
Emission lines increase in 
prominence from HD 101412 to HD 104237 and finally to
HD 190073.  Properties of the more prominent emission 
lines in these three spectra are given in 
Table~\ref{tab:emprop}.  The maximum intensity, or $I$-value, 
and equivalent
width, $W$ are with respect to the continuum; 
emission below the continuum line is not taken into
account.
\begin{table}
\caption{Properties of selected emission lines\label{tab:emprop}}
\begin{tabular}{l r r r r} \hline
               &           &           &\multicolumn{2}{c}{FWHM} \\
Star(HD)/line  & Max $I$   &   W[\AA]  &[\AA]& \kms  \\  \hline
H$\alpha$  & \\
 101412  & 3.39    &  10.5     &4.42   & 202  \\
 104237  & 3.55    &  13.3     &4.71   & 215  \\
 190073  & 6.82    &  32.2     &5.34   & 244  \\
Na D$_2$ &  \\
 101412  & 1.11    &  0.26     &2.36   & 120\\
 104237  & 1.45    &  1.03     &1.97   & 100 \\
 190073  & 1.76    &  1.38     &1.38   & 70  \\
Fe \II\ 5169 &  \\
 104237  & 1.37    &  1.04     &2.17   & 126 \\
 190073  & 1.77    &  1.89     &2.14   & 124  \\
Mg \I (b$_1$) 5184  \\
 104237  & 1.06    &  0.10    &1.48   &  86\\
 190073  & 1.45    &  1.03     &1.97   & 114\\
$[\rm O]$ \I\, 6300  & \\
 101412  & 1.07    &  0.125    &1.71   & 81  \\
 104237  & 1.04    &  0.017    &0.75   & 36  \\
 190073  & 1.11    &  0.056   &0.41   & 20 \\  \hline
\end{tabular}
\end{table}

Only a few emission lines are seen in 
HD 101412.  These include the low Balmer members, the
sodium D-lines, and 
He \I\, (D$_3$ or $\lambda$5876), which is weakly
in emission.  The strong Fe \II\, Multiplet 42, so often
in emission in young stars is in absorption.  Indeed, it is
somewhat more strongly in absorption than would be
expected from the other Fe \II\, lines, as described by
Gray and Corbally (2009, see their {\bf Nab} class).
Neither the Ca \II\, H and K lines nor the infrared triplet
is in emission.

Both HD 104237 and HD 190073 show the typical emission
lines of young stars, including the Balmer, Ca \II\,
and numerous lines from the first but mostly second 
spectra of iron-group elements.

\begin{figure}
\includegraphics[width=54mm,height=84mm,angle=270]{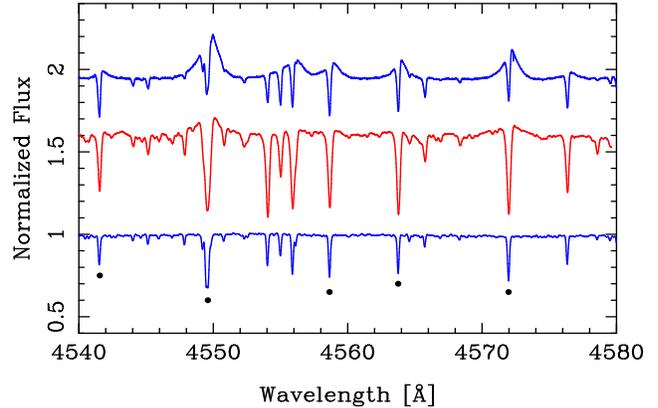}
 \caption{Emission and absorption lines in HD 104237 (center),
HD 101412 (lower), and HD 190073 (upper).  Lines marked with
filled circles are discussed in the text.
\label{fig:eml}}
\end{figure}

Figure~\ref{fig:eml} illustrates some of the weaker emission
lines in these stars. 
The strongest lines in this region are dominated by Ti \II,
$\lambda$4549.60 and 4571.97.  Weaker emissions, Fe \II\ 
$\lambda$4541.56, Cr \II\, $\lambda$4558.65, and Ti \II\,
$\lambda$4563.76 are stronger in HD 190073 than 
in HD 104237.  No emissions are evident in HD 101412.

\section{Discussion}

Herbig AeBe stars have been investigated as possible
progenitors of magnetic Ap/CP stars (Drouine et al. 
2004, Folsom, et al. 2012).  A modest fraction of 
these stars are magnetic (ca. 5-10\%  Hubrig, et al. 
2010, Alecian, et al. 2012), as is the case with their
main-sequence congeners (Power, et al. 2008).  The
chemistry of the Herbig stars is less indicative of 
an evolutionary connection.  A few of the Herbigs do
indeed have abundance excesses that could connect them
to classical, magnetic Ap stars (Vick, et al. 2011).   
However, the most
common peculiarity that has emerged so far, is a 
resemblance to the $\lambda$ Boo stars.  

In their new
detailed study, Folsom, et al. (2012) concluded that
about half of their sample of 20 stars 
showed $\lambda$ Boo
characteristics.  We had made independent
abundance studies of two of the Folsom et al. stars,
HD 101412 (Cowley, et al. 2010), and HD 190073
(Cowley \& Hubrig 2012).  The Folsom et al.
results for these stars are in excellent agreement
with ours.  These two stars as well as HD 104237
have relatively sharp spectral lines, an 
important feature for accurate analysis. 
One star, HD 101412, has
mild $\lambda$ Boo abundances, while two, HD 190073
and HD 104237 (present paper)
have solar abundances (apart from Li).  Indeed, the
connection between young stars and the $\lambda$ Boo
type has recognized for some time (Gray \& Corbally
1998).  

The prominence of $\lambda$ Boo types among the 
Herbig AeBe complicates the theoretical understanding
of these stars.  The received mechanism for Ap/CP
chemistry, diffusion (Michaud 1970), 
is quite different from that
commonly discussed for the $\lambda$ Boo stars,  gas-grain
separation (cf. discussion and references in AW).

HD 104237 has mostly solar abundances, as noted
by AW.  Its excess of
lithium is expected for a young star.  The largest 
departure from solar reported here is for zirconium,
whose excess in field stars is usually associated with
the s-process.  If there is an indication of the s-process
among related elements (Sr, Y, Ba, La), it is marginal.

\section{Acknowledgements}

We thank J. R. Fuhr J. Reader, and W. Wiese of NIST for advice on atomic data and processes, M. Briquet for help with the section on pulsations, and J. F. Gonz\'{a}lez for help with the observational material.  We are grateful S. Bagnulo and the UVESPOP team for their valuable archive. This research has made use of the SIMBAD database, operated at CDS, Strasbourg, France. Our calculations have made extensive use of the VALD atomic data base (Kupka, et al. 1999), as well as the NIST online Atomic Spectroscopy Data Center (Kramida, et al. 2013). CRC is grateful for advice and helpful conversations with many of his Michigan colleagues.  We thank the referee, B. Smalley, for valuable comments and suggestions which have strengthened this contribution.

\section*{REFERENCES}

\hangpar Acke, B., Waelkens, C. 2004, A\&A, 427, 1009 (AW)

\hangpar Aldenius, M., Lundberg, H., Blackwell-Whitehead,
R. 2009, A\&A, 502, 989

\hangpar Alecian, E., Wade. G. A., Catala, C., Grunhut, J. H., Landstreet, J. D., Bagnulo, S., et al. 2013, MNRAS, 429, 1001

\hangpar Anders, E. \& Grevesse, N. 1989, Geochim. Cosmochim.
Acta, 53, 197

\hangpar Asplund, M., Grevesse, N., Sauval, A. J., Scott, P.
2009, Ann. Rev. Astron. Ap. 47, 481

\hangpar Bagnulo, S. Jehin, E., Ledoux, C., Cabanac, R.,
Mello, R., Gilmozzi, R., et al. 2003, The Messenger,
114, 10


\hangpar B\"{o}hm, T., Catala, C., Balona, L. \& Carter,
B. 2004, A\&A, 427, 907



\hangpar Bi\'{e}mont, \'{E}., Blagoev, K., Engstrom, L.
et al. 2011, MNRAS, 414, 3350

\hangpar Blackwell, D. E., Menon, S. L. R., Petford, A. D.
\& Shallis, M. J. 1982, MNRAS, 201, 611

\hangpar Castelli, F. 2010,
wwwuser.oat.ts.astro.it/castelli/grids.html

\hangpar Castelli, F. \& Hubrig, S. 2004, A\&A, 425, 263

\hangpar Castelli, F. \& Kurucz, R. L. 2003, in
Modelling of Stellar Atmospheres, ed. N. Piskunov,
W. W. Weiss, D. F. Gray, IAU Symposium 210,
p. 424

\hangpar Cayrel, R., Faurobert-Scholl, M., Feautrier, N.,
Spielfield, A. \& Thevenin, F. 1996, A\&A, 312, 549

\hangpar Cowley, C. R., Adelman, S. J. \& Bord, D. J.
2003, in Modelling of Stellar Atmospheres, IAU Symp.
210, ed. N. Piskunov, W. W. Weiss \& D. F. Gray,
p. 261

\hangpar Cowley, C. R. \& Hubrig, S. 2012, AN, 333, 34

\hangpar Cowley, C. R., Hubrig, S., Gonz\'{a}lez \&
Savanov, I. 2010, A\&A, 523, 65



\hangpar Davidson, M. D., Snoek, L. C., Volten, H.
\& D\"{o}nszelmann, A. 1992, A\&A, 255, 457


\hangpar den Hartog, E. A., Lawler, J. E., Sobeck, J. S.,
Sneden, C. \& Cowan, J. J. 2011, ApJS, 194, 35


\hangpar Donati, J.-F., Semel, M., Carter, B. D., 
Rees, D. E. \& Collier Cameron, A. 1997, MNRAS, 291, 658.

\hangpar Dupret, M.-A, B\"{o}hm, T., Goupil, M.-J.,
Catala, C., Grigahc\`{e}ne, A. 2006, Comm. Astroseismology,
147, 72

\hangpar Dupret, M.-A., Th\'{e}ado, T, B\"{o}hm, T.,
Goupil, M.-J., Catala, C., Grigahc\`{e}ne, A. 2007,
Comm. Astroseismology, 150, 59

\hangpar Drouine, D., Wade, G. A., Bagnulo, S., 
Landstreet, J. D., Mason, E. \& Monin, D. N. 2004,
in The A-Star Puzzle, IAU Symp. 224, 506.

\hangpar Feigelson, E. D., Lawson, W. A. \& Garmire,
G. P. 2003, ApJ, 599, 1207

\hangpar Fisher, C. F. 2002, private (cf. NIST site
for Mg II)

\hangpar Folsom, C. P., Bagnulo, S., Wade, G. A., Alecian, E.,
Landstreet, J. D., Marsden, S. C. \& Wate, I. A. 2012, 
MNRAS, 422, 2072

\hangpar Fuhr, J. R. \& Wiese, W. L. 2005, in CRC 
Handbook of Chem. \& Physics, 86th ed. 10-93 (ed. D.
R. Lide, CRC Press, Boca Raton, FL)

\hangpar Fuhr, J. R. \& Wiese, W. L. 2006, J. Phys. Chem.
Ref. Data, 35, 1669

\hangpar Fumel, A. \& B\"{o}hm, T. 2012, A\&A, 540, 108 (FB)

\hangpar Grady, C. A., Woodgate, B., Torres, C. A. O.,
Henning, Th, Apai, D., Rodmann, J., et al. 2004, ApJ,
608, 809

\hangpar Gray, R. O., \& Corbally, C. J. 1998, AJ,
116, 2530

\hangpar Gray, R. O. \& Corbally, C. J. 2009, Stellar
Spectral Classification (Princeton: Series in Astrophys.),
see p. 203


\hangpar Hibbert, A., Bi\'{e}mont, \'{E}, Godefroid, M.
\& Vaeck, N. 1993, A\&AS, 99, 179





\hangpar Hubrig, S., Pogodin, M. A., Yudin, R. V.,
Sch\"{o}ller, M. \& Schnerr, R. S. 2007, A\&A, 463, 1039

\hangpar Hubrig, S., Sch\"{o}ller, M., Savanov, I.,
Gonz\'{a}lez, J. F., Cowley, C. R., Sch\"{u}tz, O.,
et al. 2010, AN, 331, 361

\hangpar Hubrig, S., Stelzer, B., Sch\"{o}ller, M.,
Grady, C., Sch\"{u}tz, Pogodin, M. A., et al. 2009,
A\&A, 502, 283




\hangpar Kelleher, D. E. \& Podobedova, L. I. 2008,
J. Phys. Chem. Ref. Data, 37, 267

\hangpar Kupka, F., Piskunov, N. E., Ryabchikova, T. A.,
Stempels, H. C., Weiss, W. W. 1999, A\&AS, 138, 119

\hangpar Kramida, A., Ralchenko, Yu., Reader, J., and NIST ASD Team (2012). NIST Atomic Spectra Database (ver. 5.0), [Online]. Available: http://physics.nist.gov/asd [2013, February 21]. National Institute of Standards and Technology, Gaithersburg, MD. 



\hangpar Kurucz, R. L. 2005, MSAIS, 8, 14

\hangpar Kurucz, R. L. 2012, http://kurucz.harvard.edu/atoms/

\hangpar Lawler, J. E., Bonvallet, G. \& Sneden, C.
2001, ApJ, 556, 452

\hangpar Lawler, J. E. \& Dakin, J. T. 1989, JOSA, 6B,
1457


\hangpar Ljung, G., Nilsson, H., Asplund, M.,
Johansson, S. 2006, A\&A, 456, 1181


\hangpar Lodders, K., Palme, H. \& Gail, H-P. 2009.
Landolt-B\"{o}rnstein, New Series, Astron. Astrophys.
(arXiv:-901.1149).

\hangpar Lyubimkov, L. S. \& Samedov, Z. A. 1987, BCrAO 77, 109 

\hangpar Mayor, M., Pepe, F., Queloz, D., et al. 2003,
Msngr, 114, 20.




\hangpar Mathur, S., Metcalfe, T. S., Woitaszek, M.,
Bruntt, H., Verner, G. A., Christensen-Dalsgaard, J.,
et al. 2012, ApJ, 749, 152

\hangpar Mel\'{e}ndez, J., Barbuy, B. 2009, A\&A,
497, 611

\hangpar Mendoza, C., Eissner, W., Le Dourneuf, M. \&
Zeippen, C. J. 1995, J. Phys. B28, 3485

\hangpar Michaud, G. 1970, ApJ, 160, 641

\hangpar Nahar, S. N. \& Pradhan, A. K. 1993, J. Phys.
B26, 1109

\hangpar Nave, G., Johansson, S., Learner, R. C. M,
Thorne, A. P. \& Brault, J. W. 1994, ApJS, 94, 221

\hangpar Nilsson, H., Ljung, G., Lundberg, H.,
Nielson, K. E. 2006, A\&A, 445, 1165

\hangpar Nitz, D. E., Kunau, A. E., Wilson, K. L. \&
Lentz, L. R. 1999, ApJS, 122, 557

\hangpar NIST 2012, http://www.nist.gov/pml/div684/grp01/data.cfm

\hangpar Ostrovskii, Yu. I. \& Penkin, N. P. 1958, Opt.
Spektrosk., 5, 345

\hangpar Pickering, J. C., Thorne, A. P., Perez, R.
2001, ApJS, 132, 403 (Erratum: ApJS, 138, 247, 2002)


\hangpar Pirronello, V. \& Strazzulla, G. 1981, A\&A,
93, 411

\hangpar Power, J., Wade, G. A., Auri\`{e}re, M, Silvester,
J. \& Hanes, D. 2008, Contrib. Astron. Obs. Skal. 
Pleso, 38, 443


\hangpar Ryabchikova T.A., Piskunov N.E., Kupka F., Weiss W.W.,
  1997, Baltic Astron., 6, 244



\hangpar Smith, V., Lambert, D. L. \& Nissen, P. E. 1998,
ApJ, 506,405

\hangpar Sobeck, J.S., Lawler, J. E., Sneden, C. 2007,
ApJ, 667, 1267

\hangpar Sugar, J. \& Corliss, C. H. 1985, J. Phys. Chem.
Ref. Data, 14, Suppl. 2

\hangpar Tachiev, G. I. \& Fisher, C. F. 2002, A\&A, 385,
716.




\hangpar Vick, M., Michaud, G., Richer, J. \& Richard, O.
2011, A\&A, 526, 37


\hangpar Wade, G. A., Bagnulo, S., Drouin, D., Landstreet, 
J. D. \& Monin, D. 2007, MNRAS, 376, 1145

\hangpar Wickliffe, M. E. \& Lawler, J. E. 1997, ApJS,
110, 163

\hangpar Yan, Z.-C \& Drake, G. W. F. 1995, Phys. Rev. 52A,
4316
\appendix

\section{Oscillator strength sources}

Keys to oscillator strengths used for the individual 
spectra are given in column 2 of Table~\ref{tab:abtab}.
A short indication is given here to complete citation
found in the references.  Compilations of NIST
(Kramida et al. 2013), VALD (Kupka, et al. 1999),
and Kurucz (2012) were used for lines not listed in
modern primary sources.

\begin{table}
\caption{Short References for Table~\ref{tab:abtab}}
\begin{tabular}{r l} \hline
Key   &  Short Ref.      \\
(1)   &Yan \& Drake (1995)         \\
(2)   &Hibbert, et al. (1993)      \\
(3)   &Tachiev \& Fisher (2002)     \\
(4)   &NIST/Kramida et al. (2013) \\
(5)   &NIST/Kramida et al. (2013) \\
(6)   &Kelleher \& Podobedova (2008) \\
(7)   &NIST/ Fisher, C. F. (2002) \\
(8)   &Mendoza et al. (1995)   \\
(9)   &Nahar \& Pradhan (1993) \\
(10)  &Kurucz (2012) \\
(11)  &VALD Ryabchikova, et al. (1997) \\
(12)  &NIST \& Aldenius, et al. (2009) \\
(13)  &Kurucz (2012)  \\
(14)  &Lawler \& Dakin (1989) \\
(15)  &Blackwell, et al. (1982) \\
(16)  &Pickering, et al. (2001) \\
(17)  &NIST/Ostrovskii \& (1958)  \\
(18)  &NIST \& Kurucz (2012)   \\
(19)  &Sobeck, et al. (2007)  \\
(20)  &Nilsson, et al. (2006),VALD LS-permitted    \\
(21)  &VALD/den Hartog et al. (2011) \\
(22)  &VALD/LS-permitted  \\
(23) &Fuhr\&Wiese (2006)   \\       
(24) &Mel\'{e}ndez\&Barbuy (2009) \\
(25) &Nitz, et al. (1999)    \\      
(26) &VALD, Wicliffe\&Lawler(1997) \\
(27) &VALD, LS-allowed \\           
(28) &NIST/Fuhr \& Wiese (2005)  \\
(29) &VALD/Kurucz (2012)  \\
(30) &Pirronello \& Strazzula (1981) \\
(31) &Bi\'{e}mont, et al. (2011) \\
(32) &Ljung, et al. (2006) \\
(33) &NIST/Davidson, et al. (1992) \\
(34) &Lawler, et al. (2001) \\
\end{tabular}
\end{table}

\section{Online material}

Equivalent widths, excitation potentials, oscillator 
strengths, and calculated abundances are available for
individual lines as online material.  

 \begin{center}
\begin{table}
\caption{Sample online material for C {\sc i}.  
Weak lines used for the adopted
abundances are marked with an asterisk}
\label{tab:online}
\begin{center}
\begin{verbatim}
Wavelength EW   Log(EW)  loggf   Chi(eV) Log(C/Ntot)  
4022.844   9.1    0.96  -2.650   7.480   -3.71*
4734.256  15.8    1.20  -2.370   7.950   -3.44*
4770.026  24.7    1.39  -2.440   7.480   -3.46*
4932.049  52.2    1.72  -1.660   7.680   -3.69
5052.167  93.4    1.97  -1.300   7.680   -3.61
6010.675   9.0    0.96  -1.940   8.640   -3.66*
7093.234  10.7    1.03  -1.700   8.650   -3.78*
7100.123  24.0    1.38  -1.470   8.640   -3.63*
7108.930  15.6    1.19  -1.590   8.640   -3.72*
7483.449  21.5    1.33  -1.370   8.770   -3.68*
7860.877  29.2    1.47  -1.150   8.850   -3.67
\end{verbatim}
\end{center}
\end{table}
\end{center}

\label{lastpage}
\end{document}